\documentclass[
 reprint,
superscriptaddress,
showpacs,preprintnumbers,
 amsmath,amssymb,
 aps,
pra,
]{revtex4-1}

\usepackage{color}
\usepackage{graphicx}
\usepackage{dcolumn}
\usepackage{bm}

\begin{document}


\title{Coherent state path integrals in the continuum}

\author{G. Kordas}
\affiliation{University of Athens, Physics Department,Panepistimiopolis, Ilissia 15771 Athens, Greece}

\author{S.I. Mistakidis}
\affiliation{Zentrum f\"{u}r Optische Quantentechnologien, Universit\"{a}t Hamburg, Luruper Chaussee 149, 22761 Hamburg,Germany}
\affiliation{University of Athens, Physics Department,Panepistimiopolis, Ilissia 15771 Athens, Greece}

\author{A.I. Karanikas}
\affiliation{University of Athens, Physics Department,Panepistimiopolis, Ilissia 15771 Athens, Greece}

\date{\today}

\begin{abstract}
We discuss the time-continuous path integration in the coherent
states basis in a way that is free from inconsistencies. Employing this notion we
reproduce known and exact results working directly in the continuum.
Such a formalism can set the basis to develop
perturbative and non-perturbative approximations already known in the quantum field
theory community. These techniques can be proven useful in a great variety of problems where
bosonic Hamiltonians are used.

\end{abstract}


\maketitle


\section{\label{sec:I}Introduction}

The widely known path integral formalism that was pioneered by
Feynman~\cite{Feynman1,Feynman2} almost seventy years ago has been
proven an extremely helpful tool for understanding and handling
quantum mechanics, quantum field theory, statistical mechanics, even
polymer physics and financial markets~\cite{Kleinert}. The
introduction of the overcomplete base of coherent
states~\cite{Klauder1,Klauder2,Klauder3,Klauder4,Weissman,Klauder5}, has
expanded the concept of path integration into a complexified phase
space enlarging its range of possible applications in many areas of
physics and chemistry, mainly as a tool for semiclassical
approximations. The path integration in terms of coherent states has
been discussed in detail in a lot of excellent
papers~\cite{Klauder1,Klauder2,Klauder3,Klauder4,Weissman,Klauder5,Xavier,Baranger,Kohetov,Stone}.
In most of them both the definition and the calculations are based
on lattice regularization and the continuum limit is taken only
after the relevant calculations have been performed. On the other
hand, quantitative differences with exact results have been
reported~\cite{Wilson} when one tries to handle coherent state path
integrals and perform calculations directly in the continuum.
A recent attempt~\cite{yanay13} to solve the
problem offers only corrections to a questionable leading term and does 
not give a definitive solution.
However, the continuum form of coherent state-based path
integration has been extensively used in quantum field theory for
perturbative approximations, e.g for resuming perturbative series or
for applying non-perturbative techniques. In this sense, it looks
annoying that the time-continuous integration in a complexified
phase space is plugged with problems.

When dealing with path integral expressions in the continuum
we have to take into account that such expressions must be
considered as formal unless a definite regularization prescription
has been given~\cite{Klauder5}. In this work we undertake the task
of establishing a time-continuous formulation of path integration in
the coherent states basis and a corresponding time-sliced
definition. In the context of the proposed formulation, the path integration
 can be performed directly in the
continuum without facing inconsistencies and reproduces the exact
results at least for the cases in which the relevant Hamiltonian is
expressed as a polynomial of creation and annihilation operators.
Such bosonic Hamiltonians are used in a great variety of important
physical problems, e.g., ultracold atoms in optical lattices~\cite{Jaks98},
cavity optomechanical systems~\cite{Aspe,Tomadin12}, non-equilibrium transport~\cite{Dujardin,Ivanov}
and other phenomena~\cite{Graefe08,Huo12,Leib10}.
So our formalism may be proven a powerful tool both for analytical
and numerical applications since it allows the use of the quantum field theory toolbox.
These techniques may be
proven helpful for extending the study of many body dynamics beyond the usual
approximate methods.

The paper is organized as follows. In Sec.\ref{sec:II} we reproduce known
results, such as the partition function for the simple case of a
harmonic oscillator, using path integration in the complexified phase
space. Then, in Sec.\ref{sec:III} we calculate the partition function for the
case of the one-site Bose-Hubbard (BH) model with time-continuous
coherent state path integrals while in Sec.\ref{sec:IV} we use this method in
order to find the exact expression for the propagator. Finally, in
Sec.\ref{sec:V} we discuss the semiclassical calculation for a Hamiltonian
that depends only on the number operator. We summarize our findings
and give an outlook in Sec.\ref{sec:VI}.

\section{\label{sec:II}A simple example}

To set the stage, we begin with the trivial case of a harmonic
oscillator
\begin{equation}
\label{eq:1}{\hat H_0} = \frac{{{{\hat p}^2}}}{2} + \frac{{{{\hat
q}^2}}}{2}.
\end{equation}
The partition function of this system, ${Z_0} = {\rm Tr}~{\rm e}^{ - \beta
{{\hat H}_0}} = \sum\limits_{n = 0}^\infty  {{\rm e}^{ - \beta \left( {n
+ 1/2} \right)}}$, can be expressed as a Feynman phase space
integral
\begin{eqnarray}
\nonumber {Z_0} &=& \int \mathcal{D} p \mathop{\int
\mathcal{D}q}\limits_{q(0)=q(\beta)} \exp \left\{ { -
\int\limits_0^\beta  {d\tau \left[ { - ip\dot q + {H_0}\left( {p,q}
\right)} \right]} } \right\}
\\
 &=& \frac{{\rm e}^{-\beta /2}}{1 - {\rm e}^{ - \beta /2}} = \sum\limits_{n = 0}^\infty  {\rm e}^{-\beta \left(n + 1/2\right)}.
\label{eq:2}
\end{eqnarray}
The integral in the left hand side (lhs) of the above expression
acquires a full meaning through its time-sliced definition. However,
in the simple case of the harmonic oscillator, the result (\ref{eq:2}) can be
derived directly in the continuum~\cite{Kleinert}. In the phase space path integral that 
appears in eq.(\ref{eq:2}) we can make the
canonical change of variables
\begin{equation}
q = \frac{1}{\sqrt 2}\left( {{z^ * } + z} \right),~~p =
\frac{i}{\sqrt 2}\left( {{z^ * } - z} \right).
\label{eq:3}
\end{equation}
In terms of these complex variables, eq.(\ref{eq:2}) is
transcribed into the following form
\begin{eqnarray}
\nonumber Z_0 &=& \mathop{\int
\mathcal{D}^2z}\limits_{{\rm periodic}} \exp \left\{ { -
\int\limits_0^\beta  {d\tau \left[ {\frac{1}{2}\left( {{z^ * }\dot z
- {{\dot z}^ * }z} \right) + {{\left| z \right|}^2}} \right]} }
\right\}\\
 &=& \sum\limits_{n = 0}^\infty  {{\rm e}^{ - \beta \left( {n + 1/2}
 \right)}}.
\label{eq:4}
\end{eqnarray}
A comment is needed at this point. In the phase space integral
 (\ref{eq:2}) the integration over $q(\tau )$ is restricted by the
periodic condition $q(0) = q(\beta )$ while the $p(\tau )$
integration is unrestricted. For the time-sliced expression that
defines the integral, this means that we are dealing with $\left(
{{q_0},...,{q_N};{q_0} = {q_N}} \right)$ ``position'' and $\left(
{{p_1},...,{p_N}} \right)$ ``momentum'' integrations. To arrive at the
periodic conditions accompanying the integral (\ref{eq:4}), one
~\cite{Kleinert} introduces a fictitious variable which is set
identically equal to ${p_N}$.

However, the partition function (\ref{eq:4}) can also be calculated by using
the coherent states basis
\begin{eqnarray}
\nonumber Z_0 &=& \int {\frac{{dzd{z^ * }}}{{2\pi
i}}} \left\langle z \right|{\rm e}^{ - \beta {{\hat H}_0}}\left| z
\right\rangle\\
 &=& {\rm e}^{ - \beta /2}\int {\frac{{dzd{z^*}}}{{2\pi i}}} \left\langle z \right|{\rm e}^{ - \beta {{\hat a}^\dag }\hat a}\left| z
 \right\rangle.
\label{eq:5}
\end{eqnarray}
Splitting the exponential into $N$ factors and using the following
resolution of the identity operator in terms of coherent states~\cite{Xavier,Baranger,Kohetov}
\begin{eqnarray}
\nonumber \hat I &=& \int {\frac{{{d^2}z}}{\pi }}
\left| z \right\rangle \left\langle z \right|: = \int {\frac{{dz~d{z^
* }}}{{2\pi i}}\left| z \right\rangle \left\langle z \right|}
\\
 &=& \int {\frac{{d{\rm Re} z~d{\rm Im} z}}{\pi }} \left| z \right\rangle \left\langle z
 \right|,
\label{eq:6}
\end{eqnarray}
we arrive at the  expression
\begin{equation}
\left\langle z \right|{\rm e}^{ - \beta {{\hat a}^\dag }\hat
a}\left| z \right\rangle  = \mathop {\lim }\limits_{N \to \infty }
\prod\limits_{j = 1}^{N - 1} {\int {\frac{{d{z_j}dz_j^ * }}{{2\pi
i}}} } {{\mathop{\rm e}\nolimits} ^{ - {f_0}\left( {{z^ * },z}
\right)}},
\label{eq:7}
\end{equation}
where the exponent has the form
\begin{eqnarray}
\nonumber {f_0}\left( {{z^ * },z} \right) &=& \sum\limits_{j = 0}^{N - 1} {\left[ {\frac{1}{2}\left( {{z_{j + 1}} - {z_j}} \right)z_{j + 1}^ *  } \right.}
\\
&& \left. { - \frac{1}{2}\left( {z_{j + 1}^ *  - z_j^ * }
\right){z_j} + \varepsilon z_{j + 1}^ * {z_j}} \right],
\label{eq:8}
\end{eqnarray}
and $\varepsilon  = \beta /N$. Note the boundary conditions in
eq.(\ref{eq:7}) that follow from the trace operation $z_N^ *  = {z^ *
},~{z_0} = z$. The integrations can be explicitly performed~\cite{Xavier,Baranger} and comparing the result with (\ref{eq:4}) we conclude that
\begin{eqnarray}
\nonumber &&\mathop{\int \mathcal{D}^2z}\limits_{{\rm periodic}} \exp \left\{ { - \int\limits_0^\beta  {d\tau \left[ {\frac{1}{2}\left( {{z^ * }\dot z - {{\dot z}^ * }z} \right) + {{\left| z \right|}^2}} \right]} } \right\}
\\
&& = {\rm e}^{ - \beta /2}\mathop {\lim }\limits_{N \to \infty } \prod\limits_{j = 0}^N {\int {\frac{{d{z_j}dz_j^ * }}{{2\pi i}}} } {{\mathop{\rm e}\nolimits} ^{ - {f_0}\left( {{z^ * },z}
 \right)}}.
\label{eq:9}
\end{eqnarray}
It is a simple exercise~\cite{Kleinert} to confirm that the factor
appearing in the right hand side (rhs) of the last equation can be absorbed into the
discretized expression by symmetrizing the time slicing of the
Hamiltonian from $z_{j + 1}^* z_j$ to $z_j^* z_j$
\begin{eqnarray}
\nonumber && \mathop{\int\mathcal{D}^2 z}\limits_{{\rm periodic}}  \exp \left\{ { - \int\limits_0^\beta  {d\tau \left[ {\frac{1}{2}\left( {{z^ * }\dot z - {{\dot z}^ * }z} \right) + {{\left| z \right|}^2}} \right]} } \right\}
\\
 &&= \mathop {\lim }\limits_{N \to \infty } \prod\limits_{j = 0}^N {\int {\frac{{d{z_j}dz_j^ * }}{{2\pi i}}} } \exp \left[ { - f_0^{(s)}\left( {{z^ * },z} \right)}
 \right],
\label{eq:10}
\end{eqnarray}
where
\begin{eqnarray}
\nonumber f_0^{(s)}\left( {{z^ * },z} \right) &=&
\sum\limits_{j = 0}^{N - 1} {\left[ {\frac{1}{2}\left(z_{j + 1} - z_j \right)z_{j + 1}^ *  } \right.}
\\
&& \left. { - \frac{1}{2}\left( {z_{j + 1}^ *  - z_j^ * }
\right){z_j} + \varepsilon z_j^ * {z_j}} \right].
\label{eq:11}
\end{eqnarray}

Despite the fact that the two sides in eq.(\ref{eq:10}) have been calculated
independently, we consider this relation as a \textit{definition} in the
sense that it gives a concrete meaning to the formal integration
over paths that go through a complexified phase space.

As a definition, eq.(\ref{eq:10}) can also be read from a different point of
view. Suppose that we are given the normal ordered Hamiltonian
$\hat{H}_1 = {\hat{a}^\dag }\hat{a}$ and we want to find the relevant
time-continuous coherent state path integral. The previous analysis
dictates that we must begin by finding the position-momentum
expression for the Hamiltonian in hand $\hat{H}_1 = {\hat{p}^2}/2 +
{\hat{q}^2}/2 - 1/2$. Then, we have to construct the Feynman phase
space path integral in which this Hamiltonian assumes its classical
version $H_1^F = {p^2}/2 + {q^2}/2 - 1/2$. Making in this
integral the variable change (\ref{eq:3}) we get $H_1^F = {\left| z
\right|^2} - 1/2$, thus obtaining the continuous path integral we
are looking for. The discretized definition of this integral can be
read from eq.(\ref{eq:10})
\begin{eqnarray}
{\rm Tr}{\rm e}^{ - \beta {{\hat H}_1}} &=&
\mathop{\int \mathcal{D}^2z}\limits_{{\rm periodic}} {{\mathop{\rm
e}\nolimits} ^{ - \int\limits_0^\beta  {d\tau \left[
{\frac{1}{2}\left( {{z^ * }\dot z - {{\dot z}^ * }z} \right) +
{H_1^F}\left( {{z^ * },z} \right)} \right]} }}
\\
\nonumber &=& {\rm e}^{\beta /2}\mathop {\lim }\limits_{N \to
\infty } \prod\limits_{j = 0}^N {\int {\frac{{d{z_j}dz_j^ * }}{{2\pi
i}}} } \exp \left[ { - f_0^{(s)}\left( {{z^ * },z} \right)} \right].
\label{eq:12}
\end{eqnarray}
In this trivial example it is useful to point out that although we
begun from a normal ordered Hamiltonian, the Hamiltonian entering
into the continuous path integral is the Weyl-symbol ${H_W}\left(
{{z^*},z} \right)$ which, in the present case, coincides with
${H_1^F}\left( {{z^ * },z} \right)$.

\section{\label{sec:III}The one-site Bose-Hubbard model}

As a less trivial example let us consider the one site BH model
\begin{equation}
{\hat H_{BH}} =  - \mu \hat n + \frac{U}{2}\hat n\left(
{\hat n - 1} \right),
\label{eq:13}
\end{equation}
where $\hat n = {\hat a^\dag }\hat a$ denotes the particle number
operator, $\mu $ is the chemical potential and $U$ the corresponding
interparticle interaction. The partition function of this system is
readily seen to be
\begin{equation}
Z_{BH} = {\rm Tr}~{\rm e}^{ - \beta {{\hat H}_{BH}}} =
\sum\limits_{n = 0}^\infty  {{\rm e}^{ - \beta \left[ { - \mu n +
\frac{U}{2}n\left( {n - 1} \right)} \right]}}.
\label{eq:14}
\end{equation}
The same result can be obtained by going directly through path
integration. As the above discussion has shown, the route begins by
using the ``position'' and ``momentum'' operators to rewrite (\ref{eq:13}) in the
form
\begin{eqnarray}
 \nonumber \hat{H}_{BH} &=&  - \frac{1}{2}\left(
{\mu  + U} \right)\left( {{{\hat p}^2} + {{\hat q}^2}}
\right)\\
 &&+ \frac{U}{8}{\left( {{{\hat p}^2} + {{\hat q}^2}} \right)^2} + \frac{\mu }{2} +
 \frac{{3U}}{8}.
\label{eq:15}
\end{eqnarray}
The partition function of the system can now be expressed as a
Feynman phase space path integral
\begin{equation}
Z_{BH} = \int \mathcal{D}p \mathop{\int
\mathcal{D}q}\limits_{q(0)=q(\beta)}
\exp\left\{-ip\dot{q}+H^F_{BH}(p,q)\right\}.
\label{eq:16}
\end{equation}
It is obvious that in the last expression, ${H_{BH}^F}$ stands for
the classical version of the quantum Hamiltonian (\ref{eq:15}). Introducing
the complex variables (\ref{eq:3}), we obtain
\begin{eqnarray}
\nonumber Z_{BH} &=& {\rm e}^{ - \beta \left(
{\frac{\mu }{2} + \frac{{3U}}{8}}
\right)}\\
\nonumber && \times \mathop{\int {\mathcal{D}^2 z}}\limits_{{\rm periodic}}  {{\mathop{\rm e}\nolimits} ^{ - \int\limits_0^\beta  {d\tau \left[ {\frac{1}{2}\left( {{z^ * }\dot z - {{\dot z}^ * }z} \right) - \left( {\mu  + U} \right){{\left| z \right|}^2} + \frac{U}{2}{{\left| z \right|}^4}} \right]}
 }}\\
&=& {\rm e}^{ - \beta \left( {\frac{\mu }{2} +
\frac{{3U}}{8}} \right)}\mathop {\lim }\limits_{N \to \infty }
\prod\limits_{j = 0}^N {\int {\frac{{d{z_j}dz_j^ * }}{{2\pi i}}} }
{{\mathop{\rm e}\nolimits} ^{ - f_{BH}^{(s)}\left( {{z^ * },z}
\right)}},
\label{eq:17}
\end{eqnarray}
where
\begin{eqnarray}
&&f_{BH}^{(s)}\left( {{z^ * },z} \right)
= \sum\limits_{j = 0}^{N - 1} {\left[ {\frac{1}{2}\left( {{z_{j +
1}} - {z_j}} \right)z_{j + 1}^ * } \right.} \label{eq:18}
\\
\nonumber &&\left. { - \frac{1}{2}\left( {z_{j + 1}^ *  - z_j^ * } \right){z_j}
- \varepsilon \left( {\mu  + U} \right){{\left| {{z_j}} \right|}^2}
+ \varepsilon \frac{U}{2}{{\left| {{z_j}} \right|}^4}} \right].
\end{eqnarray}
We shall prove that the above integral can be exactly calculated
yielding the result (\ref{eq:14}). Before this, however, a comment is in
order. The Hamiltonian entering in the last expression
\begin{equation}
H_{BH}^F\left( {{z^ * },z} \right) =  - \left( {\mu +
U} \right){\left| z \right|^2} + \frac{U}{2}{\left| z \right|^4} +
\frac{\mu }{2} + \frac{{3U}}{8},
\label{eq:19}
\end{equation}
constitutes (apart from a constant) the Weyl-symbol Hamiltonian
${H_{BH,W}}$ for the system under consideration. To understand this
point we must take a closer look at the proposed technique that
follows the route
\begin{equation}
\hat H\left( {{{\hat a}^\dag },\hat a} \right) \to \hat
H\left( {\hat q,\hat p} \right) \to {H^F}\left( {q,p} \right),
\label{eq:20}
\end{equation}
which is  a recipe for associating  an arbitrary quantum Hamiltonian
with a classical function. The key observation is that when the
quantum Hamiltonian is a polynomial in $\hat{a}$ and ${\hat{a}^\dag }$
the respective time-slicing of the Feynman path integrals~\cite{Kleinert} leads to expressions that differ from the Wigner
transformation
\begin{equation}
{H_W}\left( {p,q} \right) = \int\limits_{ - \infty
}^\infty  ds~{\rm e}^{ips} \left\langle {q - \frac{s}{2}} \right|\hat
H\left| {q + \frac{s}{2}} \right\rangle,
\label{eq:21}
\end{equation}
which defines the Weyl-symbol, by at most a constant.

The calculation of the integral (\ref{eq:17}) proceeds with the use of a
Hubbard-Stratonovich~\cite{Stratonovich,Hubbard,Muhlshegel,Halpern,Jevicki} transformation. This can
be realized by the introduction of the collective field $\zeta  =
{\left| z \right|^2}$ and the use of the functional identities

\begin{eqnarray}
\nonumber 1 &=& \int \mathcal{D}\zeta \delta \left[ {\zeta  - {{\left| z \right|}^2}} \right],\\
\delta \left[\zeta  - \left| z \right|^2 \right] &=&
\int \mathcal{D}\sigma {\rm e}^{ - i\int\limits_0^\beta  {d\tau \sigma \left(\zeta - \left| z \right|^2 \right)} }.
\label{eq:22}
\end{eqnarray}
In this way the integral under consideration takes the form
\begin{eqnarray}
\nonumber Z_{BH} &=& {\rm e}^{ - \beta \left(
{\frac{\mu }{2} + \frac{{3U}}{8}} \right)}\int \mathcal{D}\zeta \int
\mathcal{D}\sigma
\\
\nonumber &&\times {\rm e}^{ - i\int\limits_0^\beta  {d\tau \sigma \zeta  - \frac{U}{2}\int\limits_0^\beta  {d\tau {\zeta ^2} + \left( {\mu  + U} \right)\int\limits_0^\beta  {d\tau } \zeta } }}
\\
&&\times \mathop{\int \mathcal{D}^2 z}\limits_{{\rm periodic}}
{{\mathop{\rm e}\nolimits} ^{ - \int\limits_0^\beta  {d\tau \left[
{\frac{1}{2}\left( {{z^ * }\dot z - {{\dot z}^ * }z} \right) -
i\sigma {{\left| z \right|}^2}} \right]} }}.
\label{eq:23}
\end{eqnarray}
Here, the last functional integration can be performed directly in
the continuum~\cite{Kleinert}. The result reads as follows
\begin{equation}
\mathop{\int\mathcal{D}^2 z}\limits_{{\rm periodic}}
{\rm e}^{ - \int\limits_0^\beta  {d\tau \left[
{\frac{1}{2}\left(z^*\dot z - \dot{z}^* z\right) -
i\sigma {{\left| z \right|}^2}} \right]} } =
\frac{{\rm e}^{\frac{i}{2}\int\limits_0^\beta  {d\tau \sigma } }}{{1 -
{\rm e}^{i\int\limits_0^\beta  {d\tau \sigma } }}}.
\label{eq:24}
\end{equation}

Inserting this into eq.(\ref{eq:23}), and assuming that a small positive
imaginary part accompanies the field $\sigma $, we can immediately find that
\begin{eqnarray}
\nonumber Z_{BH} &=& {\rm e}^{ - \beta \left({\frac{\mu }{2} + \frac{{3U}}{8}} \right)}\int \mathcal{D}\zeta \int\mathcal{D}\sigma
\\
\nonumber&&\times {\rm e}^{ - i\int\limits_0^\beta  {d\tau \sigma \zeta  - \frac{U}{2}\int\limits_0^\beta  {d\tau {\zeta ^2} + \left( {U + \mu } \right)\int\limits_0^\beta  {d\tau \zeta } } } }\sum\limits_{n = 0}^\infty  {{\rm e}^{i\left( {n + \frac{1}{2}} \right)\int\limits_0^\beta  {d\tau \sigma } }}
 \\
\nonumber&=& {\rm e}^{ -\beta \left( {\frac{\mu }{2} + \frac{{3U}}{8}} \right)}\int \mathcal{D}\zeta {\rm e}^{ - \frac{U}{2}\int\limits_0^\beta  {d\tau {\zeta ^2} + \left( {U + \mu } \right)\int\limits_0^\beta  {d\tau \zeta } }
}\\
&&\times \sum\limits_{n = 0}^\infty  \int \mathcal{D}\sigma {\rm e}^{ - i\int\limits_0^\beta  d\tau \sigma \left(\zeta - n - 1/2\right) }.
\label{eq:25}
\end{eqnarray}
The integration over the field $\sigma $ results to a functional
delta function that enforces the field $\zeta $ to be a constant:
$\zeta  = n + 1/2$. Thus we get
\begin{eqnarray}
\nonumber Z_{BH} &=& {\rm e}^{ - \beta
\frac{{3U}}{8}}\sum\limits_{n = 0}^\infty {\rm e}^{ - \frac{U}{2}\beta
\left(n + \frac{1}{2}\right)^2 + U\beta \left( {n + \frac{1}{2}} \right) + \mu \beta n}
\\
 &=& \sum\limits_{n = 0}^\infty {\rm e}^{ -\beta \left[ - \mu n + \frac{U}{2}n\left( {n - 1} \right)\right]}.
\label{eq:26}
\end{eqnarray}
Before proceeding, a comment is needed. Let us suppose that one
tries to calculate the integral (\ref{eq:17}) by using polar coordinates $z =
\sqrt{r} {\rm e}^{i\theta }$~\cite{Wilson}. In this case the continuum action is supposed to have the form
\begin{eqnarray}
&&\int\limits_0^\beta  {d\tau \left[ {ir\dot \theta  - \left( {\mu  + U} \right)r + \frac{U}{2}{r^2}} \right]}  = ir\left( \beta  \right)\theta \left( \beta  \right)\\
\nonumber &&  - ir\left( 0 \right)\theta \left( 0 \right) + \int\limits_0^\beta  {d\tau \left[ { - i\dot r\theta  - \left( {\mu  + U} \right)r + \frac{U}{2}{r^2}}
 \right]}.
\label{eq:27}
\end{eqnarray}
The measure of the functional integration is taken to be
\begin{equation}
\int\mathcal{D}^2z  = \int \mathcal{D}r \mathcal{D}\theta
= \mathop {\lim }\limits_{N \to \infty } \prod\limits_{j = 0}^N
\int\limits_0^\infty d{r_j}\int\limits_0^{2\pi } \frac{d\theta_j}{2\pi }.
\label{eq:28}
\end{equation}
The integral over $\theta $  ensures that $r$ is a constant and the
first term in the lhs of eq.(\ref{eq:27}) enforces this constant to be an
integer. In this manner, one arrives at the wrong conclusion that
\begin{equation}
Z_{BH} = {\rm e}^{ - \beta \left( \frac{\mu }{2} +
\frac{{3U}}{8}\right)}\sum\limits_{n = 0}^\infty  {\rm e}^{-\left(\mu + U \right)n\beta - \frac{U}{2}{n^2}\beta}.
\label{eq:29}
\end{equation}
The problem has nothing to do with the BH path integral
(\ref{eq:17}); It persists even for the trivial case of the simple harmonic
oscillator (\ref{eq:4}) and the well-known result (\ref{eq:14}) is not reproduced. 
The culprit for these wrong results is the fact that the parameter $\theta(t)$, being the phase of $z(t)$,
is a multivalued function: At every instant $t$ it is possible to add an arbitrary integer multiple of $2\pi$ without changing
${\rm e}^{i\theta(t)}$. Thus the use of Leibnitz rule that led to the
expression (\ref{eq:27}) was completely illegal~\cite{kleinert14}. The problem
persists even in the discrete version of the relevant integral: A calculation based on the use
of polar coordinates fails to reproduce the correct continuum limit. The proper way to take
into account the periodicity of $z$ is by writing~\cite{Kleinert}
\begin{equation}
z\left(\tau\right) = \frac{1}{\sqrt{\beta}}\sum\limits_{m = - \infty }^\infty  z_m {\rm e}^{ -i\frac{{2\pi m}}{\beta }\tau }.
\label{eq:31}
\end{equation}
In this way, the correct results emerge in both the continuum and
the discretized versions of the path integral.

\section{\label{sec:IV}Correlation Functions}

As long as we are interested in the partition function of a system,
the measure of integration in terms of the $\left( {p,q} \right)$
variables can be immediately translated into the measure in terms of
the $\left( {z,{z^ * }} \right)$  variables. The situation changes
when we are interested in calculating path integrals with specific
boundary conditions in the complexified phase space.  This kind of
calculations is tightly related with correlation functions that are
the basic tools needed in any actual calculation pertaining to
systems with interactions.

We can express propagators in the coherent state language beginning
with the definition
\begin{eqnarray}
\nonumber &&\left\langle z_b \right|\hat{U}\left( {T,0} \right)\left|z_a \right\rangle =\\
&&\mathop{\int \mathcal{D}^2z}\limits_{\substack{z^*(T)=z^*_b \\ z(0)=z_a}} {\rm e}^{ - \Gamma
_{ba}}{\rm e}^{i\int_0^T {dt\left[
{\frac{i}{2}\left(z^*\dot{z} - \dot{z}^* z \right) -
{H^F}\left(z^*,z\right)} \right]} }.
\label{eq:32}
\end{eqnarray}
In this expression we have denoted the time evolution operator as
\begin{equation}
\hat{U}\left(t_b,t_a\right) = \hat T\exp \left\{- i\int\limits_0^T {dt\hat H\left( t \right)} \right\},
\label{eq:33}
\end{equation}
and we have used the abbreviation
\begin{equation}
\Gamma _{ba} = \frac{1}{2}\left( \left| z_b \right|^2 + \left| z_a \right|^2 \right) -
\frac{1}{2}\left( z_b^* z\left( T \right) + z^*\left( 0 \right)z_a \right).
\label{eq:34}
\end{equation}
The interpretation of (\ref{eq:32}) is the following: In the lhs
one begins by dividing the time interval $\left( {T,0} \right)$ into
small pieces  $\varepsilon  = T/N$, inserting in each step the
coherent state resolution of the identity operator and following the
standard~\cite{Xavier,Baranger} procedure is led to the symmetric
time-sliced version of the coherent state path integral. The limit
$N \to \infty $ of this discretized expression defines the path
integral that appears in the rhs in eq.(\ref{eq:32}).

The consequences of the definition (\ref{eq:32}) can be trivially checked in
the case of a harmonic oscillator with a frequency $\omega $.
Starting from the right hand side we solve the classical equations
of motion with the boundary conditions $z_{cl.}^* (T) = z_b^*,~z_{cl.}(0) = z_a$ finding that
\begin{equation}
z_{cl.} = z_a{\rm e}^{i\omega t},~z_{cl.}^* = z_b^* {\rm e}^{ - i\omega \left( {T - t} \right)}.
\label{eq:35}
\end{equation}
Then we perform the replacements $z \to z + {z_{cl}}$ and $z^* \to {z^*} + z_{cl.}^* $ in order to find
\begin{eqnarray}
\nonumber&&\int\limits_{z_b^*,z_a} \mathcal{D}^2 z
{\rm e}^{ - {\Gamma _{ab}}}{{\mathop{\rm e}\nolimits} ^{i\int\limits_0^T
{dt\left[ {\frac{i}{2}\left( {{z^ * }\dot z - {{\dot z}^ * }z}
\right) + \omega {{\left| z \right|}^2}} \right]} }}
\\
\nonumber &=&\exp \left\{ z_b^* {z_a}{\rm e}^{i\omega T} - \frac{1}{2}\left(
\left| {{z_b}} \right|^2 + \left| {{z_a}} \right|^2\right) \right\} \times
\\
&&\times \mathop{\int\mathcal{D}^2z}\limits_{z_b^*  = {z_a} = 0}  {\rm e}^{i\int\limits_0^T {dt\left[
{\frac{i}{2}\left( {{z^* }\dot z - \dot{z}^* z} \right) + \omega {{\left| z \right|}^2}} \right]} }.
\label{eq:36}
\end{eqnarray}
According to (\ref{eq:32}) the functional integral in the rhs of
eq.(\ref{eq:36}) is the vacuum expectation value of the time evolution
operator of the harmonic oscillator
\begin{equation}
\left\langle 0 \right|\hat{U}\left( {T,0} \right)\left| 0 \right\rangle  = {\rm e}^{ - i\omega T/2}.
\label{eq:37}
\end{equation}
Inserting eq.(\ref{eq:37}) into eq.(\ref{eq:36}) we can derive the harmonic oscillator
propagator in the coherent state representation. This result could
also have been produced~\cite{Xavier,Baranger} directly from the lhs of the
definition (\ref{eq:32}).

Another simple case in which the definition (\ref{eq:32}) can be used for
calculations directly in the continuum is the case of the
BH model (\ref{eq:13}). In this framework, the propagator
\begin{equation}
{K_{ba}} = \left\langle {{z_b}} \right|{\rm e}^{ - iT{{\hat
H}_{BH}}}\left| {{z_a}} \right\rangle,
\label{eq:38}
\end{equation}
is immediately seen to have the form
\begin{eqnarray}
 K_{ba} &=& \sum\limits_{n,m} \langle z_b | n\rangle \langle n| {\rm e}^{ - iT\hat{H}_{BH}} |m \rangle \langle m | z_a\rangle
 \\
\nonumber&=& {\rm e}^{- \frac{1}{2}\left( \left| z_b \right|^2 + \left|z_a\right|^2 \right)}
 \sum\limits_n \frac{\left(z_b^* z_a \right)^n}{n!} {\rm e}^{iT\mu n - \frac{iU}{2}n\left(n -1\right)}.
\label{eq:39}
\end{eqnarray}
Then, using the identity
\begin{eqnarray}
{\rm e}^{-i\frac{{TU}}{2}n\left(n - 1\right)} &=& {\rm e}^{i\frac{UT}{8}} {\rm e}^{-i\frac{TU}{2} \left(n - 1/2\right)^2} \label{eq:40}\\
\nonumber &=& {\rm e}^{i\frac{UT}{8}} \sqrt{\frac{T}{2\pi iU}} \int\limits_{-\infty }^\infty  d\omega {\rm e}^{i\frac{T}{2U} \omega^2
+ iT\omega \left(n - 1/2\right)},
\end{eqnarray}
we can rewrite the propagator into the following exact form~\cite{Wilson}
\begin{eqnarray}
\nonumber K_{ba} &=& {\rm e}^{i\frac{UT}{8}}
\sqrt{\frac{T}{{2\pi iU}}} \int\limits_{-\infty }^\infty d\omega
\exp \left\{ {i\frac{T}{{2U}}{\omega^2} - \frac{{i\omega T}}{2}} \right.
\\
&& + \left. z_b^* {z_a}{\rm e}^{i\left( \omega  + \mu \right)T}
 - \frac{1}{2}\left(\left| z_b \right|^2 + \left| z_a \right|^2 \right) \right\}.
\label{eq:41}
\end{eqnarray}
We can arrive at the same result starting from the functional integral
\begin{equation}
K_{ba} = \mathop{\int
\mathcal{D}^2z}\limits_{\substack{z^*(T) = z_b^* \\ z(0) = z_a}} {\rm e}^{ -
{\Gamma _{ba}}}{{\rm e} ^{i\int_0^T
{dt\left[ {\frac{i}{2}\left( {{z^* }\dot z - \dot{z}^* z}
\right) - H_{BH}^F\left(z^*,z\right)} \right]} }},
\label{eq:42}
\end{equation}
in which the Hamiltonian has already be defined in eq.(\ref{eq:19}).

Once again, the Hubbard-Stratonovich transformation can be used to
recast the integral (\ref{eq:42}) into the following form
\begin{eqnarray}
\nonumber K_{ba} &=& {\rm e}^{ - iT\left( {\frac{\mu}{2} + \frac{{3U}}{8}} \right)} {\rm e}^{ - \frac{1}{2}\left( {{{\left|
z_b \right|}^2} + \left| z_a \right|^2} \right)} \int \mathcal{D}\zeta \int \mathcal{D}\sigma
\\
&&\times {\rm e}^{-i\int\limits_0^T dt\sigma \zeta  - i\frac{U}{2}\int\limits_0^T dt \zeta ^2
+ i\left(\mu + U\right)\int\limits_0^T dt\zeta}{\tilde{K}_{ba}},
\label{eq:43}
\end{eqnarray}
where the kernel reads
\begin{eqnarray}
\nonumber \tilde{K}_{ba} &=& \mathop{\int
\mathcal{D}^2z}\limits_{z_b^*,z_a} {\rm e}^{i\int\limits_0^T dt\left[ {\frac{i}{2}\left( {{z^* }\dot z -
{\dot{z}^* }z} \right) + \sigma {{\left| z \right|}^2}} \right] + \frac{1}{2}\left(z_b^* z\left( T \right) + z^*\left(0\right)z_a\right)}
\\
&=& \exp \left\{ {\frac{i}{2}\int\limits_0^T {dt\sigma }
+ z_b^* {z_a}{\rm e}^{i\int\limits_0^T {dt\sigma } }} \right\}.
\label{eq:44}
\end{eqnarray}
Note that in order to arrive at the result indicated in the second
line of the above expression, we have made the replacements $z \to z + z_{cl.}$
and $z^* \to z^* + z_{cl.}^*$ where
\begin{equation}
z_{cl.} = {z_a}{\rm e}^{-\frac{i}{\hbar}\int\limits_0^t dt'\sigma},~z_{cl.}^*  = z_b^ * {\rm e}^{-\frac{i}{\hbar
}\int\limits_t^T dt'\sigma }
\label{eq:45}
\end{equation}
are the solutions of the classical equations of motion, and at the
same time we have used the vacuum expectation value of a harmonic
oscillator with a time-depended frequency~\cite{Kleinert}. In
order to proceed further we expand the second term that appears in
the exponential factor (\ref{eq:44}) and insert the result into eq.(\ref{eq:43}) where
the integration over $\sigma $ yields the constraint $\zeta  = n +
1/2$. Thus the propagator now reads
\begin{eqnarray}
\nonumber K_{ba} &=& {\rm e}^{-iT\frac{{3U}}{8}}{\rm e}^{-\frac{1}{2}\left(\left|z_b\right|^2 + \left|z_a\right|^2\right)}
\\
\nonumber &&\times \sum\limits_{n = 0}^\infty  {\frac{\left(z_b^* z_a \right)^n}{{n!}}{\rm e}^{iT\mu n - i\frac{U}{2}T{{\left(n + \frac{1}{2}\right)}^2} + iUT\left(n + \frac{1}{2}\right)}}
 \\
\nonumber &=& {\rm e}^{-iT\frac{{3U}}{8}} {\rm e}^{-\frac{1}{2}\left(\left|z_b\right|^2 + \left|z_a\right|^2\right)}\int\limits_{-\infty }^\infty
dx{\rm e}^{-i\frac{UT}{2} x^2 + iTUx}\\
&&\times \sum\limits_{n = 0}^\infty \frac{\left(z_b^* z_a{\rm e}^{iT\mu}\right)^n}{n!} \delta \left(x - n - \frac{1}{2}\right).
\label{eq:46}
\end{eqnarray}
Moreover, by inserting into this expression the identity
\begin{equation}
\delta \left( {x - n - 1/2} \right) =
T\int\limits_{ - \infty }^\infty  {\frac{{d\omega }}{{2\pi }}} {\rm e}^{
- i\omega T\left( {x - n - 1/2} \right)},
\label{eq:47}
\end{equation}
we arrive at the exact result
\begin{eqnarray}
\nonumber K_{ba} &=& {\rm e}^{-iT\frac{3U}{8}}{\rm e}^{-\frac{1}{2}\left(\left|z_b\right|^2 + \left|z_a\right|^2\right)}
\\
\nonumber &&\times \frac{T}{2\pi} \int\limits_{-\infty }^\infty  d\omega {\rm e}^{\frac{i\omega T}{2}
+ z_b^* z_a {\rm e}^{iT\left(\mu  + \omega\right)} }
\\
\nonumber && \times \int\limits_{-\infty}^\infty dx {\rm e}^{-i\frac{UT}{2} x^2 + iTUx - i\omega Tx}
\\
\nonumber &=& {\rm e}^{i\frac{UT}{8}}{\rm e}^{-\frac{1}{2}\left(\left|z_b\right|^2 + \left|z_a\right|^2\right)}
\\
&&\times \sqrt{\frac{T}{2\pi iU}} \int\limits_{-\infty}^\infty \omega {\rm e}^{i\frac{T}{2U} \omega^2 - \frac{i\omega T}{2} + z_b^* {z_a}{\rm e}^{i\left(\omega + \mu\right)T}}.
\label{eq:48}
\end{eqnarray}

\section{\label{sec:V}Semi-classical calculations}

To probe in a transparent way the classical limit, one can introduce~\cite{Kohetov,Wilson} the dimensionless parameter $h$  through the
rescaling $\left( {\hat a,{{\hat a}^\dag }} \right) \to \left( {\hat
a,{{\hat a}^\dag }} \right)/\sqrt h $. In this notation $\left[
{\hat a,{{\hat a}^\dag }} \right] = h$  and $\left| z \right\rangle
= {\rm e}^{ - {\left| z \right|^2}/2h} \sum\limits_{n = 0}^\infty
{\frac{{{{\left( {z/h} \right)}^n}}}{{\sqrt {n!} }}} \left| n
\right\rangle $  while the classical limit is achieved at the limit
$h \to 0$. The quantum BH Hamiltonian (\ref{eq:13}) is written as
${\hat H_{BH}} =  - \mu \hat n + \frac{U}{2}\hat n\left( {\hat n -
h} \right)$ and the exact propagator (\ref{eq:48}) assumes the form
\begin{equation}
K_{ba} = {\rm e}^{i\frac{hUT}{8}}\sqrt {\frac{T}{{2\pi ihU}}}
\int\limits_{-\infty }^\infty  {d\omega } \exp \left\{
{\frac{1}{h}{\Phi_\omega } - \frac{{i\omega T}}{2}} \right\},
\label{eq:49}
\end{equation}
where
\begin{equation}
\Phi_\omega = i\frac{T}{2U}{\omega^2} + z_b^*
{z_a}{\rm e}^{i\left( {\omega  + \mu } \right)T} - \frac{1}{2}\left(
\left|z_b \right|^2 + \left|z_a\right|^2 \right).
\label{eq:50}
\end{equation}
At the limit $h \to 0$ the integral (\ref{eq:49}) can be evaluated
~\cite{Wilson} by finding the stationary points of ${\Phi_\omega }$
. At the same result one can arrive starting from the path integral
(\ref{eq:42}) expressed in terms of the rescaled variables
\begin{eqnarray}
K_{ba} &=& \mathop{\int\mathcal{D}^2z}\limits_{\substack{z^*(T) = z_b^ */\sqrt h \\ z(0) = {z_a}/\sqrt h}} {\rm e}^{ - {\Gamma
_{ba}}/h} \exp \left\{ \phantom{\int\limits_0^T}\right. \label{eq:51}\\
\nonumber && \left.  {\frac{i}{h}\int\limits_0^T {dt\left[ {\frac{i}{2}\left( {{z^ * }\dot z - {{\dot z}^ * }z} \right) - H_{BH}^F\left( {{{\left| z \right|}^2};h} \right)} \right]} }
 \right\}.
\end{eqnarray}
In the above integral the Hamiltonian is the rescaled version of the
function appearing in eq.(\ref{eq:19})
\begin{eqnarray}
\nonumber H_{BH}^F\left( {{{\left| z
\right|}^2};h} \right) &=&  - \left( {\mu + hU} \right)h{\left| z
\right|^2}\\
&& + {h^2}\frac{U}{2}{\left| z \right|^4} + \frac{h\mu}{2} +
 {h^2}\frac{3U}{8}.
\label{eq:52}
\end{eqnarray}
We shall consider here the case of an arbitrary Hamiltonian as long
as it has the form $\hat H = H\left( {\hat n} \right)$.  In this
case $H^F = H^F\left( {{{\left| z \right|}^2};h} \right)$ and
the correlation function (\ref{eq:43}) can be written as follows
\begin{eqnarray}
K_{ba} &=& {\rm e}^{ - \frac{1}{{2h}}\left( {\left|z_b\right|^2 + \left|z_a\right|^2} \right)} \label{eq:53}\\
\nonumber && \times \int\mathcal{D}\zeta \int\mathcal{D}\sigma
{\rm e}^{ - \frac{i}{h}\int\limits_0^T dt\sigma \zeta  - \frac{i}{h}\int\limits_0^T dt H^F \left(\zeta ;h\right)}
\tilde{K}_{ba}\left( h \right).
\end{eqnarray}
The factor ${\tilde K_{ba}}$  in the last expression is the rescaled
version of eq.(\ref{eq:44})
\begin{equation}
{\tilde K_{ba}}\left( h \right) = \exp \left\{
\frac{i}{2h}\int\limits_0^T dt\sigma + \frac{z_b^* z_a}{h}{\rm e}^{\frac{i}{h}\int\limits_0^T dt\sigma} \right\}.
\label{eq:54}
\end{equation}
Inserting eq.(\ref{eq:54}) into eq.(\ref{eq:53}) and repeating the steps of the
previous section we arrive at the following result
\begin{eqnarray}
\nonumber K_{ba} &=& {\rm e}^{-\frac{1}{{2h}}\left(
{{{\left| {{z_b}} \right|}^2} + {{\left| {{z_a}} \right|}^2}}
\right)}
\\
&&\times \sum\limits_{n = 0}^\infty  {\frac{{{{\left( {z_b^
* {z_a}/h} \right)}^n}}}{{n!}}} {\rm e}^{ - \frac{i}{h}T H^F\left( {n +
\frac{1}{2};h}
 \right)}.
\label{eq:55}
\end{eqnarray}
This expression can be compared with the standard semiclassical
analysis where one has to solve the classical equations ${\dot
z_{cl.}} =  - i\partial {H_F}/\partial z_{cl.}^ * $ and $\dot
z_{cl.}^ *  = i\partial {H_F}/\partial {z_{cl.}}$ with boundary
conditions ${z_{cl.}}\left( 0 \right) = {z_a}$ and $z_{cl.}^ *
\left( T \right) = z_b^* $ respectively. This task can be
considerably simplified by the introduction of an auxiliary field
$\sigma$ that serves as a functional Lagrange multiplier
\begin{equation}
{H^F} \to {H^F}\left( \zeta  \right) + \frac{\sigma
}{h}\left( {\zeta  - {{\left| z \right|}^2}} \right).
\label{eq:56}
\end{equation}
In this treatment, which is obviously equivalent to the
Hubbard-Stratonovich transformation we adopted in the previous
sections, the result indicated in eq.(\ref{eq:54}) coincides with the careful
calculation of the fluctuation determinant presented in
~\cite{Wilson}. Minimizing the final result with respect to $\sigma$ one
arrives at the expression indicated in eq.(\ref{eq:55}).

Supposing that the Hamiltonian we are dealing with is an analytic
function of $h$ we write
\begin{eqnarray}
\nonumber H^F\left(\zeta ;h\right) &=& \sum\limits_{k = 0}^N {{h^k}H_F^{(k)}} \left(\zeta ;0\right),
\\
H_F^{(k)}\left( {\zeta ;0} \right) &=& \frac{1}{{k!}}\frac{{{\partial
^k}}}{{\partial {h^k}}}{\left. {{H^F}\left( {\zeta ;h} \right)}
\right|_{h = 0}}.
\label{eq:57}
\end{eqnarray}
If the original Hamiltonian was a polynomial in powers of ${\left| z
\right|^2}$  the highest power appearing in each term
$H_F^{(k)}\left( \zeta  \right)$ is ${\zeta ^k}$. Thus
\begin{eqnarray}
\nonumber H^F\left( {\zeta ;h} \right) &=&
h\left( {{a_1}\zeta  + {a_0}} \right) + {h^2}\left( {{b_2}{\zeta ^2} + {b_1}\zeta  + {b_0}} \right)
\\
&& + \sum\limits_{k = 3}^N {{h^k}H_F^{(k)}} \left( {\zeta ;0}
 \right).
\label{eq:58}
\end{eqnarray}
In this equation we wrote
\begin{eqnarray}
\nonumber H_F^{(1)}\left( {0;\zeta } \right) &=& a_1 \zeta + a_0,
\\
H_F^{(2)}\left(0;\zeta\right) &=& {b_2}{\zeta ^2} + {b_1}\zeta + b_0.
\label{eq:59}
\end{eqnarray}
Neglecting the last term in eq.(\ref{eq:58}) and following the algebra
presented in the previous section we can find
\begin{eqnarray}
\nonumber {K_{ba}} &=& {\rm e}^{ - iT\left( {{a_0} + {a_1}/2} \right) - iTh\left( {{b_0} - b_1^2/4{b_2}}
\right)}\\
&&\times \sqrt {\frac{T}{{4\pi ih{b_2}}}} \int\limits_{ -
\infty }^\infty d\omega {\rm e}^{\frac{1}{h}{\Phi _\omega } +
\frac{{i\omega T}}{2}\left( {1 + {b_1}/{b_2}} \right)}.
\label{eq:60}
\end{eqnarray}
Here
\begin{equation}
\Phi_\omega = i\frac{T}{{4{b_2}}}{\omega^2} + z_b^* z_a{\rm e}^{i\left(\omega - a_1 \right)T} - \frac{1}{2}\left(
\left|z_b\right|^2 + \left|z_a\right|^2
\right).
\label{eq:61}
\end{equation}
In the BH model the relevant parameters are
\begin{eqnarray}
\nonumber a_0 &=& \mu /2,~a_1 = - \mu\\
b_0 &=& 3U/8,~b_1 = - U,~b_2 = U/2.
\label{eq:62}
\end{eqnarray}
and the result (\ref{eq:48}) is retrieved. It is obvious that having
neglected the higher order terms in the expansion (\ref{eq:58}) the integral
in eq.(\ref{eq:60}) must be evaluated in terms of the stationary points of
the function (\ref{eq:61}).

\section{\label{sec:VI}Conclusions and outlook}

Second quantized Hamiltonians for bosonic systems are used in a
great variety of physical problems~\cite{Jaks98,Tomadin12,Aspe,Dujardin,Ivanov,Graefe08,Huo12,Leib10}. Furthermore, the
experimental advances call for theoretical methods that will
allow the study of the many-body dynamics deep in the quantum regime.
In the present work we have introduced a method for defining and
handling time continuous coherent state path integrals without
facing inconsistencies. Such a path integral formalism
opens, in principle at least, new possibilities for the analytical
study of a variety of second quantized models.
The aim of this paper is not the presentation of new
results. It is, rather, a try to set a solid basis for really
interesting calculations which can go beyond the already known
approximate methods.

In our approach, the Hamiltonian that weights the paths in
the complexified phase space is produced through three simple steps.
In the first step one rewrites the second quantized Hamiltonian
$\hat H\left( \hat{a},\hat{a}^\dag\right)$ in terms of
``position'' and ``momentum'' operators. The second step consists of
constructing the Feynman phase-space integral in which the
classical form of this Hamiltonian $H^F\left( {p,q} \right)$
enters. The third step is just a canonical change of variables that
produces the final form $H^F\left(z,z^*\right)$ which
enters into the time-continuous form of the coherent state path
integral. We have followed this simple method for the case of the
one-site BH model and we have derived the correct
expressions for the partition function and the propagator of the
system. We have also discussed a semiclassical calculation
pertaining to a Hamiltonian that depends only on the number
operator.

In a forthcoming study, we intend to use non-perturbative techniques,
already known in the quantum field theory community, in order to
study the dynamics in realistic systems and compare our results with that of already
known approximate methods.
Another direction would be the combination of
our technique with the so-called
Feynman-Vernon influence functional formalism for studying open
many-body systems, like the open Bose-Hubbard chains or cavity systems~\cite{Brenn13}, which are of
increasing interest both theoretically~\cite{Witt11,Kor12}
and experimentally~\cite{Bakr10,Baron13,Ender13}.

{}

\end{document}